\documentclass[10pt,final,conference,letterpaper,romanappendices,twocolumn]{IEEEtran}
\usepackage{cite}
\usepackage{epsfig}
\usepackage{enumerate}
\usepackage{amsthm}
\usepackage{amsmath,amssymb,amsfonts,amstext,amsbsy,amsopn,dsfont}
\usepackage{cases}
\usepackage{sublabel}
\usepackage{array}

\newtheorem{proposition}{\textbf{Proposition}}

\newcommand{\defn}{\triangleq}
\newcommand{\dif}{\textmd{d}}

\begin{document}

\title{Adaptive Downlink CoMP in Heterogeneous Cellular Networks with Imperfect Overhead Messaging}

\author{Chun-Hung Liu \\Department of Electrical and Computer Engineering \\National Chiao Tung University\\ Hsinchu, Taiwan}

\maketitle

\begin{abstract}
Coordinated multi-point (CoMP) transmission is an effective means of improving network throughput in heterogeneous cellular networks (HetNets). However, its performance is seriously weakened if imperfect coordination happens between base stations (BSs). Many prior CoMP works do not consider inter-cell overhead message delays such that a seemingly astonishing CoMP throughput gain is attained. In this paper, the quantization error and delay that actually exist in overhead messages was modeled and we developed a much tractable SIR model based on the stochastic geometry framework. We proposed adaptive CoMP that is applied to downlink zero-forcing beamforming (ZFBF) and it can mitigate the interference from the coordinated cells with delayed overhead messages. The bounds on the complementary cumulative distribution function (CCDF) of the SIR of a user are characterized such that the average throughput of a user is able to be analytically evaluated. Numerical results show that the proposed adaptive CoMP scheme can make the throughput gain very robust to the overhead delay and thus significantly increase the throughput even when BSs are not perfectly coordinated.   
\end{abstract}

\section{Introduction}
As we have witnessed the rapid evolution of wireless cellular technologies, boosting the transmission rate of users is the most salient feature in each generation of wireless cellular systems. No matter which cellular generation we are approaching, the common hurdle we will face is to achieve high throughput demands in the danger of bandwidth deficit. Increasing the signal-to-interference ratio (SIR) is an effective means of elevating data rate in an interference-limited wireless cellular network. In a heterogeneous cellular network (HetNet), (end) users can certainly have a better SIR provided that interferences from different types of base station (BS) can be effectively mitigated. Therefore, coordination or cooperation techniques between BSs is one of the approaches to attaining the high throughput goal. Recently, coordinated multi-point (CoMP) transmission has attracted much attention since it can make multiple BSs coordinate/cooperate to improve network throughput if perfect and timely coordination between BSs is always possible\cite{Gesbert10CoMPSummary}. However, when CoMP has imperfect inter-cell overhead messaging, throughput could be significantly degraded and how to mitigate this problem is still not addressed too much until now.  

In the early work on CoMP schemes, the issues of imperfect inter-cell overhead messaging, for example, quantization error and delay, are completely overlooked\cite{SomSha00,HuaVen04,CadJaf08}. Although perfect coordination messaging facilitates the analysis of the CoMP fundamental performance, in fact those analytical results could be far away from the practically observed scenario\cite{Qcom12ITA}. Some recent work already investigated CoMP with practical overhead messaging. For example, references \cite{SanSomPooSha09,huang2013joint} looked at the capacity limit problem of inter-cell overhead channels for CoMP joint processing with shared user data. Although quantization inaccuracy in overhead messaging can be characterized by the limited feedback model, the overhead delay problem is either ignored or characterized by an oversimplified delay model\cite{JindalRVQ,Ako10Correlated}. 

In addition to the issue of imperfectly modeling overhead messaging in the previous works on CoMP, another issue in the previous works is the BS distribution models that are based on either the grid model or the Wyner model. These two BS models have been shown to be an inaccurate model for HetNets \cite{AndBacGan10,DhiGanBacAnd11}. A more appropriate model for HetNets is based on the stochastic geometry framework that models the locations of BSs in HetNets as one or more planar Poisson Point Processes (PPPs) since multiple independent PPPs are able to characterize the random locations of  different types of BSs. Another advantage of modeling a HetNet by PPPs is the tractability of deriving the SIR distribution that is used to evaluate the CoMP throughput since the SIR expression should include the effects of imperfect overhead messaging that significantly depend on which models are adopted to characterize the overhead impairments in backhaul links.   

Our recent work \cite{XiaLiuAndrews13} justifies that quantization error and delay in overhead messages significantly weaken the downlink CoMP throughput gain. However, it does not provide an effective method to mitigate interferences from the coordinated cells with out-of-dated overhead messages. In this paper, we propose an adaptive downlink CoMP scheme for the coordinated cells which do not receive the updated overhead messages in time. In this scheme, the coordinated cells which receive updated overhead messages within a predesignated waiting time window schedule a transmission; otherwise they do not. With this adaptive CoMP scheme, the bounds on the CCDF of the user's SIR  are found and they can be used to characterize the average throughput of the user. Numerical results show that the adaptive CoMP scheme has the capability of remarkably mitigating the loss of the CoMP throughput gain due to delayed overhead messaging.

\section{System Model and Preliminaries}\label{Sec:SysModDefns}
Suppose an interference-limited HetNet consists of $K$ different types of base stations (e.g. macrocells, picocells, femtocells, etc.) and each specific type is referred as a tier, which is named a $K$-tier HetNet. BSs in the $k$th tier form a homogeneous PPP $\Phi_k$ of intensity $\lambda_k$ on the plane $\mathbb{R}^2$, and they have the same transmit power $p_k$, number of antennas $n_k$, path loss exponent $\alpha_k$. These network parameters are usually distinct in different tiers since different types of BSs are designed and installed based on different supporting needs.  For instance, femtocells typically have much lower transmit power and fewer antennas if compared with macrocells. To facility the analysis in the following,  a reference user equipped with a single antenna is assumed to be located at the origin. Let  $\mathtt{B}_{i,k}$ denote the \textit{$i$th closest BS to the reference user in tier $k$} and its location.

The cell association algorithm is performed by users that measure the long-term average powers of the downlink pilot signals from different BSs. A user will associate with the BS that can provide it the strongest average power. In other words, the (reference) user will associate with $\mathtt{B}_{1,k_*}$ with subscript $k_*$ given by 
\begin{align}\label{Eqn:OptBSTier}
k_* = \arg \max_{k=1,2,\ldots, K} \{P_k |\mathtt{B}_{1,k}|^{-\alpha_k}\},
\end{align}
which excludes fading effects since they are supposed to be averaged out in a long term sense. For downlink CoMP ZFBF,  $\mathtt{B}_{1,k_*}$ is able to mitigate/cancel its interference from  some coordinated BSs. Note that different BSs could cancel interferences from \textit{different }numbers of other cells.  

\textbf{The coordination set of $\mathtt{B}_{1,k_*}$}. 
Let $\mathcal{S}_{1,k_*}$ denote  the coordination set of $\mathtt{B}_{1,k_*}$.  In this paper, we will not specify how to select $\mathcal{S}_{1,k_*}$ and assume it is determined in advance since our interest is  in the impact of realistic inter-cell overhead signaling on the downlink CoMP ZFBF in a selected set $\mathcal{S}_{1,k_*}$. To coordinate all BSs in $\mathcal{S}_{1,k_*}$,  the coordinated parameters, such as channel states and scheduling information of users, that are dependent on various CoMP schemes need to be timely updated for each BS in $\mathcal{S}_{1,k_*}$. These parameters are of time-varying nature due to user mobility and/or dynamic network environments.

\textbf{Duration Modeling of Overhead Messages}. 
The coordinated parameters are assumed to remain unchanged in a time duration before they are updated. Let $L_{i,k}$ denote their lifetime duration and assume all lifetime durations are i.i.d. random variables with a Gamma distribution with parameter $m$, i.e. $L_{i,k}\sim\Gamma\left(m,\frac{1}{m\mu_{i,k}}\right)$ where $\mu_{i,k}=1/\mathbb{E}[L_{i,k}]$ does not depend on $m$. Using the Gamma distribution with parameter $m$ to characterize the statistics of $L_{i,k}$ is much more general since it covers several different distributions (e.g. $m=1$ for an exponential distribution with rate $1/\mu_{i,k}$,  $m= \infty$ corresponding to a deterministic distribution, etc.) Thus the coordinated parameters vary in different time slots, and once they change an overhead message will be generated and sent to its all coordinated BSs so that those BSs can adopt an appropriate coordination strategy accordingly. Overhead messages should be generated under a sufficient frequency that does not cause too much burden in the backhual link.  Note that each overhead message only has a lifetime of $L_{i,k}$ since  a new overhead message is generated  after $L_{i,k}$.

 \begin{figure}[!t]
 \centerline{
 \includegraphics[width=3.5in,height=1.35 in]{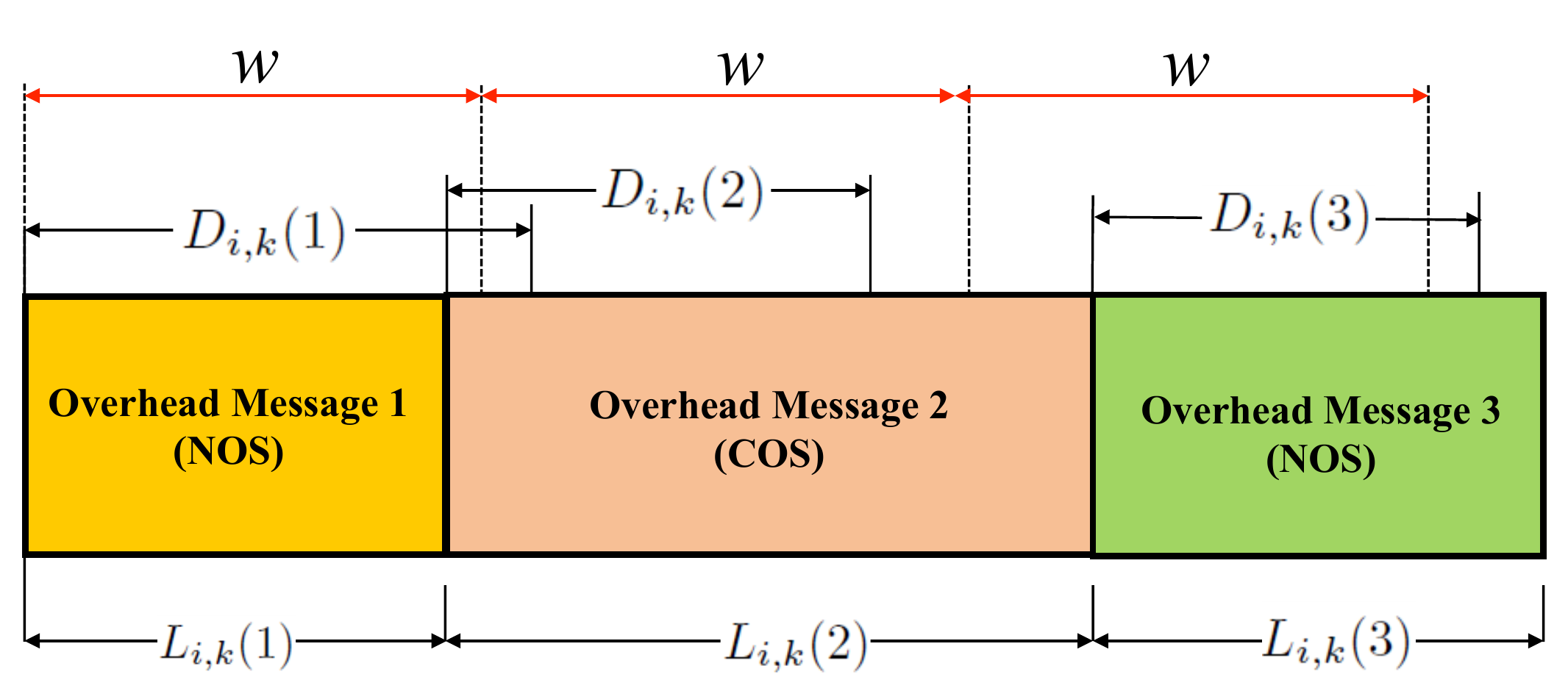}}
 \caption{Coordinated overhead  state (COS) and non-coordinated overhead state (NOS) of a coordinated BS $\mathtt{B}_{i,k}$. $D_{i,k}(j)$ and $L_{i,k}(j)$ are the overhead delay and lifetime durations for overhead message $j$, respectively. Although the waiting time duration help  $\mathtt{B}_{i,k}$ decide its coordination state, it may also make $\mathtt{B}_{i,k}$ be in the wrong state. For example, $\mathtt{B}_{i,k}$ is declared in NOS, but it could be in COS if $w$ were large enough.} \label{PicCoMPSystem}
 \end{figure} 

Here we would like to characterize how delayed overhead messages impact the performance of downlink CoMP ZFBF if all  coordinated cells use zero-forcing precoders to null their mutual interferences. Therefore, the coordinated parameter for BS $\mathtt{B}_{i,k}$ is the normalized channel state information (CSI) given by
$\mathbf{h}_{i,k}\triangleq\frac{\mathbf{H}_{i,k}}{\|\mathbf{H}_{i,k}\|}$ in which $\mathbf{H}_{i,k}$ is an $n_k \times 1$ fading channel gain vector from $\mathtt{B}_{i,k}$ to the user and each component of $\mathbf{H}_{i,k}$ is i.i.d. complex Gaussian $\mathcal{CN}(0,1)$ assuming all channels undergo uncorrelated Rayleigh fading. Note that the coherent time of $\mathbf{H}_{i,k}$ can be viewed as $L_{i,k}$, i.e. the lifetime duration of $\mathbf{H}_{i,k}$.  The user uses $\mathbf{h}_{i,k}$ to search through a codebook $\mathcal{C}_{i,k}$ known by itself and $\mathtt{B}_{i,k}$, which consists of $2^{b_{i,k}}$ codewords. The user will select the codeword $\mathbf{c}_{i,k}$ that maximizes $|\mathbf{c}_{i,k}\mathbf{h}_{i,k}|$.  Then the $b_{i,k}$-bit index of the codeword $\mathbf{c}_{i,k}$ is sent back to $\mathtt{B}_{1,k_*}$ and $\mathtt{B}_{1,k_*}$ then  forwards the  overhead message to $\mathtt{B}_{i,k}$ over a backhaul link. Once $\mathtt{B}_{i,k}$ receives the overhead message,  it can choose a zero-forcing precoding vector $\mathbf{f}_{i,k}$ such that $|\mathbf{c}_{i,k}\mathbf{f}_{i,k}|\approx 0$.  

\section{SIR Characterization of Adaptive CoMP}
 There are two pivotal issues in practical overhead messaging: transmission delay and quantization error. Transmission delay is incurred by propagation time, traffic congestion and/or hardware delay in the backhual link. According to the magnitude of the delay, a coordinated cell can be in either the coordinated overhead state (COS) or non-coordinated overhead state (NOS). Only the interferences from the coordinated cells with a COS can be mitigated at the user. In other words, the user's SIR can be increased if the interferences from the coordinated cells with a NOS can be reduced. However, in fact all coordinated cells are unable to know in advance  to which coordinated states they belong so that the coordinated cells with a NOS could seriously undermine the user's SIR.   See Fig. \ref{PicCoMPSystem} for an illustration example.
 
 These two coordination states seriously impact the user's received signal-to-interference (SIR) since  the interferences from $\mathtt{B}_{i,k}$ in these two states are quiet different. In the NOS, the channel  $\mathbf{H}_{i,k}$ has already changed but $\mathtt{B}_{i,k}$ does not know this and thus the zero-forcing precoder $\mathbf{f}_{i,k}$ cannot be updated. Thus, statistically $|\mathbf{f}_{i,k}\mathbf{H}_{i,k}|^2$ is not reduced under this situation. Whereas if $\mathtt{B}_{i,k}$ receives the new overhead message in time (i.e. it is in the COS) then it can adjust its zero-forcing precoder $\mathbf{f}_{i,k}$ accordingly so that $\mathbf{f}_{i,k}$ minimizes the interference $|\mathbf{f}_{i,k}\mathbf{H}_{i,k}|^2$\cite{JindalRVQ,Loveetal08}. In order to reduce the interferences from the coordinated cells with a NOS, we propose an adaptive CoMP scheme that makes all the coordinated cells decide in which state they are according to whether or not they can receive an updated overhead message within a predesignated waiting duration.  Let $w$ be the predesignated waiting time duration for all the coordinated BSs in $\mathcal{S}_{1,k_*}$.  $\mathtt{B}_{i,k}\in\mathcal{S}_{1,k_*}$  will not transmit if it does not receive an updated overhead message within $w$.   
 
Another issue existing in overhead messaging is the messaging quantization error that is induced by the finite quantization bits $b_{i,k}$. The exact impact of $b_{i,k}$ depends on the specific CoMP scheme and the overhead codebook $\mathcal{C}_{i,k}$ (See \cite{Loveetal08} for an overview). Here we are interested in its impact in CoMP ZFBF. Let $\gamma_{1,k_*}$ be the user's SIR in CoMP ZFBF and it can be expressed as 
\begin{align} \label{EquGeneralSIR}
    \gamma_{1,k_*}=\frac{p_{k_*} G_{1,k_*} |\mathtt{B}_{1,k_*}|^{-\alpha_{k_*}} }{ \sum_{\mathtt{B}_{i,k}\in \bigcup\limits_{k=1}^K \Phi_k \setminus \mathtt{B}_{1,k_*}}\delta_{i,k}p_k  G_{i,k}|\mathtt{B}_{i,k}|^{-\alpha_k}  },
\end{align} 
where $G_{1,k_*}\triangleq |\mathbf{f}_{1,k_*}\mathbf{H}_{1,k_*}|^2\sim \chi^2 (2n_{k_*}-2|\mathcal{S}_{1,k_*}|)$, $|\mathcal{A}|$ is the cardinality of set $\mathcal{A}$, $G_{i,k}\triangleq |\mathbf{f}_{1,k}\mathbf{H}_{1,k}|^2\sim  \chi_2^2$, $\delta_{i,k}$ is the \emph{interference-coordinated  scaling factor} for $\mathtt{B}_{i,k}$ and given by
\begin{align} \label{Eqn:delta}
\delta_{i,k} &= \mathds{1}(\mathtt{B}_{i,k} \notin \mathcal{S}_{1,k_*})+0\cdot\mathds{1}(\{\mathtt{B}_{i,k} \in \mathcal{S}_{1,k_*}\} \cap \{w<D_{i,k}\})\nonumber\\
&+2^{-\frac{b_{i,k}}{n_k-1}} \mathds{1}(\{\mathtt{B}_{i,k} \in \mathcal{S}_{1,k_*}\}\cap\{\min\{L_{i,k}, w\}\geq D_{i,k}\})\nonumber\\
&+\mathds{1}(\{\mathtt{B}_{i,k} \in \mathcal{S}_{1,k_*}\}\cap \{w \geq D_{i,k}\geq L_{i,k}\}),
\end{align}
where $D_{i,k}$ is the transmission delay from BS $\mathtt{B}_{1,k_*}$ to $\mathtt{B}_{i,k}\in\mathcal{S}_{1,k_*}$ and $\mathds{1}(A)$ is an indicator function which is unity if event $A$ is true, otherwise zero. The value of $\delta_{i,k}$ depends on the coordination status of a BS $\mathtt{B}_{i,k}$. For example, if $\mathtt{B}_{i,k}\in\mathcal{S}_{1,k_*}$ and $w, L_{i,k}\geq D_{i,k}$ (i.e. the coordinated BS $\mathtt{B}_{i,k}$ receives the updated overhead message in time), the interference from   $\mathtt{B}_{i,k}$ can be reduced by $2^{-\frac{b_{i,k}}{n_k-1}}$ fold. Also, the distributions of random variables $G_{1,k_*}$ and $G_{i,k}$ are elaborated  respectively in the following:\\
(i) The serving BS $\mathtt{B}_{1,k_*}$ needs to choose its precoder $\mathbf{f}_{1,k_*}$ such that $G_{1,k_*} \sim \chi_{2n_{k_*}-2|\mathcal{S}_{1,k_*}|}^2$ since it has to maximize $G_{1,k_*}$ as well as null the interferences  from the coordinated BSs in $\mathcal{S}_{1,k_*}$ \cite{Jindal11Adhoc}.\\
(ii) A coordinated BS $\mathtt{B}_{i,k} \in \mathcal{S}_{1,k_*}$ cannot null its interference in the NOS since its zero-forcing precoder $\mathbf{f}_{i,k}$ is independent of $\mathbf{H}_{i,k}$ and thus $G_{i,k} \sim  \chi_2^2$. In the COS, $\mathtt{B}_{i,k}$ receives the updated overhead that is used to choose $\mathbf{f}_{i,k}=\arg\min\{|\mathbf{c}_{i,k}\mathbf{f}_{i,k}|^2\}$. Usually,  $G_{i,k}$ would not be zero and dependent on the overhead codebook $\mathcal{C}_{i,k}$. In this paper, a random vector quantization (RVQ) codebook is used for $\mathcal{C}_{i,k}$, which is commonly used in previous CoMP ZFBF\cite{JindalRVQ,Loveetal08,Jindal11Adhoc}. With an RVQ-based codebook $\mathcal{C}_{i,k}$, we can have $G_{i,k} \sim 2^{-\frac{b_{i,k}}{n_k-1}}$ for $\mathtt{B}_{i,k}$ in the COS. \\

\section{Throughput Analysis of Adaptive CoMP ZFBF}
In this section, the throughput  analysis for adaptive CoMP with overhead delay will be presented. First, we study the  the complementary cumulative distribution function (CCDF) of the user's SIR in
CoMP ZFBF since we will need it to propose a throughput evaluation framework for the adaptive CoMP with overhead delay. Although the exact result of the CCDF of the SIR is analytically intractable,  we can find its bounds as shown in the following proposition.
\begin{proposition}\label{Prop:BoundsCCDFSIR}
Suppose the coordinated set $\mathcal{S}_{1,k_*}$ of BS $\mathtt{B}_{1,k_*}$ is given and the number of the BSs with a COS in $\mathcal{S}_{1,k_*}$ is $m$. The bounds on the CCDF of the user's SIR parameterized by $m$ are given in \eqref{Eqn:BoundKtierSIR} 
\begin{figure*}[!t]
\begin{equation}\label{Eqn:BoundKtierSIR}
F^c_{\gamma_{1,k_*}}(\beta; m)
\begin{cases} 
\geq 
1-\frac{\beta\Gamma\left(1+ \frac{\alpha_{k_*}}{2} \right)}{n_{k_*}-|\mathcal{S}_{1,k_*}|-1} \left[\sum\limits_{i=2}^{\infty} \mathbb{E}[ \delta_{i,k_*}] \frac{(i-1)!}{\Gamma\left(i+\frac{\alpha_{k_*}}{2}\right)}+ \sum\limits_{\begin{subarray}{c}
k=1\\
k \neq k_*
\end{subarray}}^K \frac{P_k}{P_{k_*}} \sum\limits_{i=1}^{\infty} \mathbb{E}[\delta_{i,k}] \frac{(\lambda_{k_*} \pi)^{-\frac{\alpha_{k_*}}{2}}(i-1)!}{(\lambda_k \pi)^{\frac{-\alpha_k}{2}}\Gamma\left(i+\frac{\alpha_k}{2}\right)}\right]\\
\leq \exp\left[-\left(\pi\tilde{\lambda}_*\right)^{1-\frac{\alpha_{k_*}}{\alpha_{\max}}}\Gamma\left(1+\frac{2}{\alpha_{\max}}\right)\left(\frac{\beta[3^{-\alpha_{\max}}\delta_{k_*}]+(2m+3)^{-\alpha_{\max}}(1-\delta_{k_*})]}{(n_{k_*}-|\mathcal{S}_{1,k_*}|) \Gamma \left(1- \frac{\alpha_{k_*}}{2}\right)}\right)^{\frac{2}{\alpha_{\max}}}\right]\\
\end{cases}
\end{equation}
\normalsize \hrulefill
\end{figure*}
where  $\tilde{\lambda}_*= \sum_{k=1}^{K} \lambda_k \left(p_k/p_{k_*}\right)^{\frac{2}{\alpha_k}}$, $\alpha_{\max}\triangleq \max\{\alpha_1,\ldots,\alpha_K\}$, $\delta_{k}\defn\min_i \mathbb{E}[\delta_{i,k}]$, $|\mathcal{S}_{1,k_*}|<n_{k_*}$ is the cardinality of $\mathcal{S}_{1,k_*}$, and 
\begin{align}
\mathbb{E}[\delta_{i,k}] =& \mathds{1}[\mathtt{B}_{i,k}\notin \mathcal{S}_{1,k_*}]+2^{-\frac{b_{i,k}}{n_k-1}}\mathds{1}[\mathtt{B}_{i,k}\in \mathcal{S}_{1,k_*}]\cdot\nonumber\\
& \left\{F^c_{L_{i,k}}(w)F_{D_{i,k}}(w)+F_{L_{i,k}}(w)-\mathbb{E}[F_{L_{i,k}}(D_{i,k})] \right\}\nonumber\\
&+\mathds{1}[\mathtt{B}_{i,k}\in \mathcal{S}_{1,k_*}]\left\{F_{D_{i,k}}(w)-\mathbb{E}[F_{D_{i,k}}(L_{i,k})]\right\}.\label{Eqn:AvgDelta}
\end{align}
\end{proposition}    
\begin{IEEEproof}
See Appendix \ref{App:ProofCCDFSIR}
\end{IEEEproof}

The bounds in \eqref{Eqn:BoundKtierSIR} both are the decreasing function of the average interference-reduced factor $\mathbb{E}[\delta_{i,k}]$. Since  $\mathbb{E}[\delta_{i,k}]$ highly depends on $w$, we should choose an appropriate value of $w$ to reduce $\mathbb{E}[\delta_{i,k}]$  as much as possible. The coordination status of a BS in $\mathcal{S}_{1,k_*}$ is also affected by $w$ very much as shown in the following proposition.
\begin{proposition} \label{Prop:TimeFrac}
 The time fraction of the COS for  a BS $\mathtt{B}_{i,k}\in\mathcal{S}_{1,k_*}$ performing the adaptive CoMP ZFBF scheme is
 \begin{align}\label{Eqn:CoorNatTime}
 \eta_{i,k}=&\mu_{i,k}\bigg\{\int_{0}^{w}F^c_{L_{i,k}}(x)\dif x -F^c_{L_{i,k}}(w)\bigg(wF_{D_{i,k}}(w)\nonumber\\
 &-\int_{0}^{w}F_{D_{i,k}}(x)\dif x\bigg)
 +F_{L_{i,k}}(w)\mathbb{E}\bigg[L_{i,k}F_{D_{i,k}}(L_{i,k})\nonumber\\
 &-\int_{0}^{L_{i,k}}F_{D_{i,k}}(x)\dif x\bigg] \bigg\}
 \end{align}
where $F_Z(\cdot)$ and $F^c_Z(\cdot)$ are the CDF and CCDF of random variable $Z$, respectively.
\end{proposition}
\begin{IEEEproof}
See Appendix \ref{App:TimeFracCOS}
\end{IEEEproof} 

\begin{figure}[!t]
\centering
\includegraphics[width=3.75in, height=2.2in]{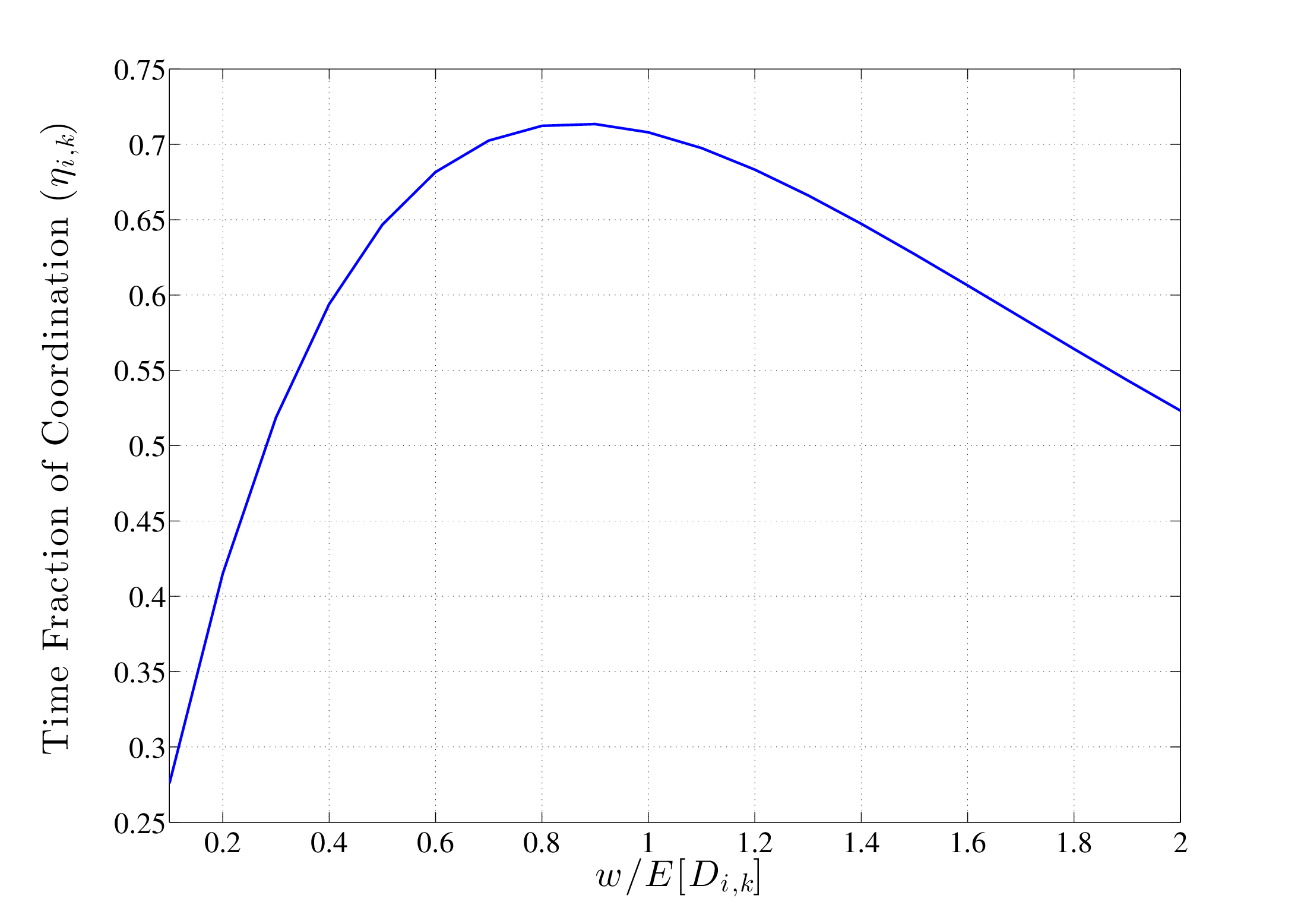}
\caption{A simulation example of $\eta_{i,k}$ for $L_{i,k}\sim \Gamma(1,80\text{ms})$ and $D_{i,k}\sim$  uniformly distribution over [0,150 ms].}
\label{Fig:TimeFrac}
\end{figure}

Prior work does not consider the overhead delay and the transmission decisions of BSs based on it. Thus, the time fraction $\eta_{i,k}$ in \eqref{Eqn:CoorNatTime} can be viewed a metric of evaluating how CoMP is effectively performed since it represents how much probability that a coordinated BS really has the COS in a long-term sense. Also, we have to note that the higher $\eta_{i,k}$, the more user's throughput benefited by CoMP. That means, it is important to determine an appropriate value of $w$ that could maximize $\eta_{i,k}$ once the distribution of $D_{i,k}$ is specified. Fig.  \ref{Fig:TimeFrac} shows a simulation example of $\eta_{i,k}$ for $L_{i,k}\sim \Gamma(1,80\text{ ms})$ and $D_{i,k}$ is uniformly distributed in [0,150 ms]. As can be seen, the optimal value of $w$ that maximizes $\eta_{i,k}$  is about $0.83\mathbb{E}[D_{i,k}]$. Using too large or too small $w$ significantly reduces the time fraction of the COS of a BS. In fact, a better method of finding an optimal $w$ is to solve the following optimization problem of $w$:
\begin{equation}
\min_{w>0} \frac{\mathbb{E}[\delta_{i,k}(w)]}{\eta_{i,k}(w)}, 
\end{equation}
which can be done numerically if the distributions of $L_{i,k}$ and $D_{i,k}$ are designated. 

The following proposition provides a metric of the average of the user's throughput under the adaptive CoMP scheme.  
\begin{proposition} \label{Prop:UserThroughput}
For the adaptive CoMP ZFBF scheme without user data sharing, the ergodic throughput of the reference user in a $K$-tier HetNet is 
\begin{align} \label{Eqn:AvgThroughput}
\mathcal{T}_{1,k_*} = \sum_{v=0}^{|\mathcal{S}_{1,k_*}|} \eta^v_{i,k_*}(1-\eta_{i,k_*})^{|\mathcal{S}_{1,k_*}|-v}\int\limits_{0}^{\infty}\frac{F^c_{\gamma_{1,k_*}(v)}(x;m)}{\ln 2(x+1)}\dif x.
\end{align}

\end{proposition}
\begin{IEEEproof}
Let $V$ be the random number of the BSs in $\mathcal{S}_{i,k_*}$ that are in COS. Since we know that each BS in $\mathcal{S}_{1,k_*}$ is either in NOS or in COS, $V$ is a binomial random variable, i.e. 
$$\mathbb{P}[V=v]=\eta^v_{i,k_*}(1-\eta_{i,k_*})^{|\mathcal{S}_{1,k_*}|-v}.$$
Therefore, the ergodic throughput per unit bandwidth of a user can be expressed as
\begin{align*}
	\mathcal{T}_{1,k_*} & =\sum_{v=0}^{|\mathcal{S}_{1,k_*}|}\mathbb{P}[V=v]\mathbb{E}[\log_2(1+\gamma_{1,k_*})]                                                                         \\
	                    & = \sum_{v=0}^{|\mathcal{S}_{1,k_*}|} \eta^v_{1,k_*}(1-\eta_{1,k_*})^{|\mathcal{S}_{1,k_*}|-v}\int\limits_0^{\infty} \frac{\mathbb{P}[\gamma_{1,k_*}(v)\geq x]}{\ln 2(x+1)}\dif x,
\end{align*}
and the result in \eqref{Eqn:AvgThroughput} is readily acquired. 
\end{IEEEproof}

As shown in \eqref{Eqn:AvgThroughput}, the CoMP throughput is apparently impacted by the overhead delay through the time fraction $\eta_{i,k_*}$. It is also affected by the finite bit size of an overhead message since the CCDF of the SIR of the user is affected by the overhead quantization error, for example, as shown in \eqref{Eqn:delta} for CoMP ZFBF. Thus, the throughput evaluation formula in \eqref{Eqn:AvgThroughput} is a more realistic metric since it indeed reflects the throughput reduction due to imperfect overhead messaging. Although the closed form result of $\mathcal{T}_{i,k_*}$ cannot be obtained, its bounds can be promptly acquired by substituting \eqref{Eqn:BoundKtierSIR} into \eqref{Eqn:AvgThroughput}.

\section{Numerical Results}
In this section, we simulate how much the average throughput of a user can be achieved under the adaptive CoMP ZFBF scheme with imperfect overhead signaling. Suppose there are three tiers in the HetNet, i.e. $K=3$. For simplicity, the first tier consists of macro BSs and the serving BS is in this tier (i.e. $\mathtt{B}_{1,k_*}=\mathtt{B}_{1,1}$), and  1 coordinated BS in $\mathcal{S}_{1,1}$ that causes the strongest interference to the user. Other simulation parameters are listed as follows: The numbers of antennas, the path loss exponents, transmit powers and densities for the three tiers are $n_1=8$, $n_2=4$, $n_3=2$, $\alpha_1=4$, $\alpha_2=3.5$, $\alpha_3=3$, $p_1=40$W, $p_2=5$W, $p_3=0.5$W, and $L_{i,k}\sim\Gamma(1, 80\text{ ms})$.    

\begin{figure}[!t]
\centering
\includegraphics[width=3.75in, height=2.2in]{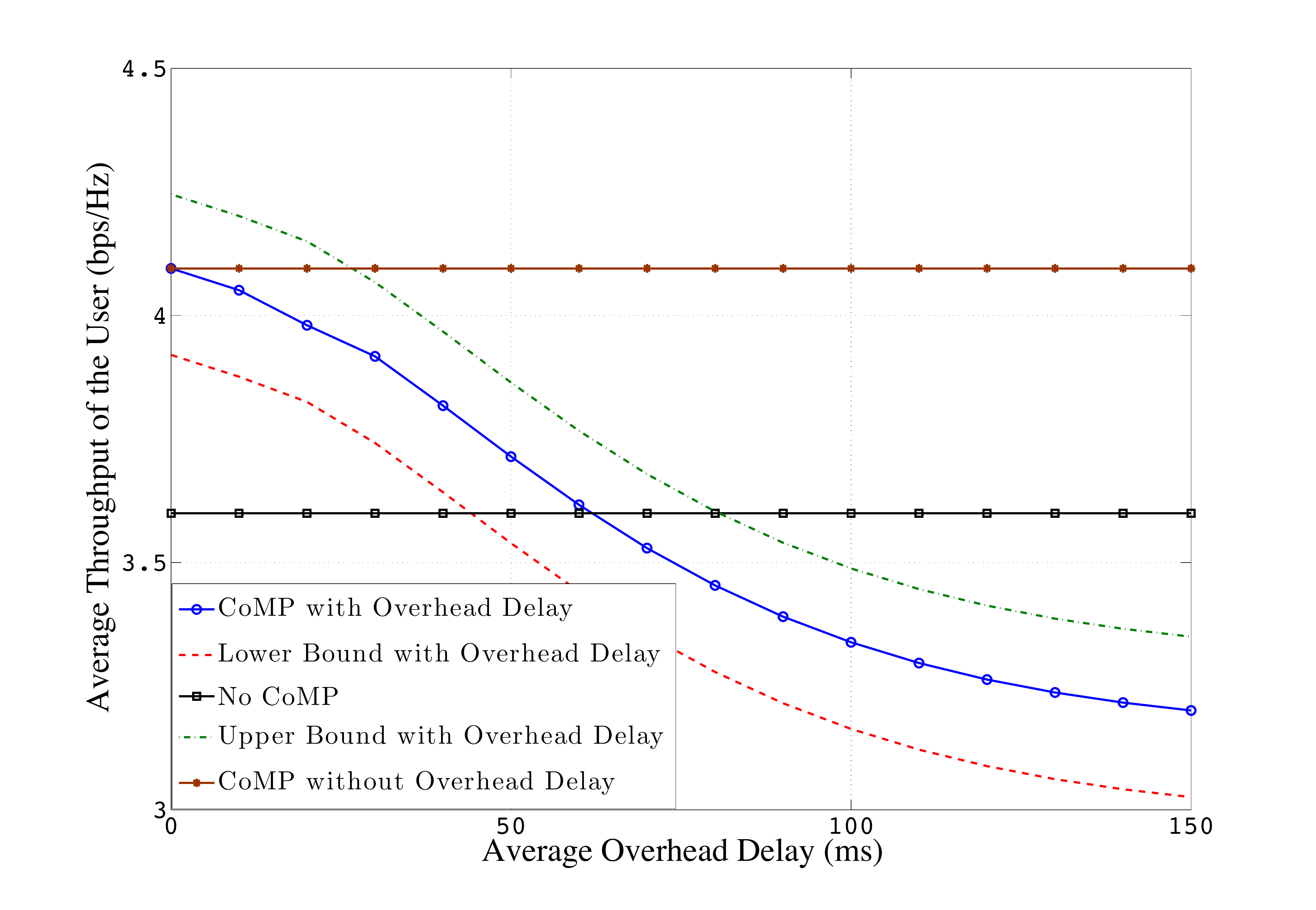}
\caption{User's average throughput vs. the average overhead channel delay for CoMP ZFBF without adaptation. The overhead bit size is $b_{i,k} = 3(n_k-1)$, which gives $2^{-b_{i,k}/(n_k-1)}=0.125$.}.
\label{Fig:ThrputCoMP}
\end{figure}

First, consider the simulation case of CoMP ZFBF without adaptation (i.e. $w=\infty$). Fig. \ref{Fig:ThrputCoMP} shows the numerical results of the user's average throughput versus the average overhead delay $\mathbb{E}[D_{1,1}]$. As expected, the user throughput dramatically decreases as the average overhead delay increases and  CoMP ZFBF does not provide any throughput gain if the average delay is over about 60ms. This points out that CoMP even reduces the user's throughput if too much overhead delay happens in the backhual links. 

\begin{figure}[!t]
\centering
\includegraphics[width=3.75in, height=2.2in]{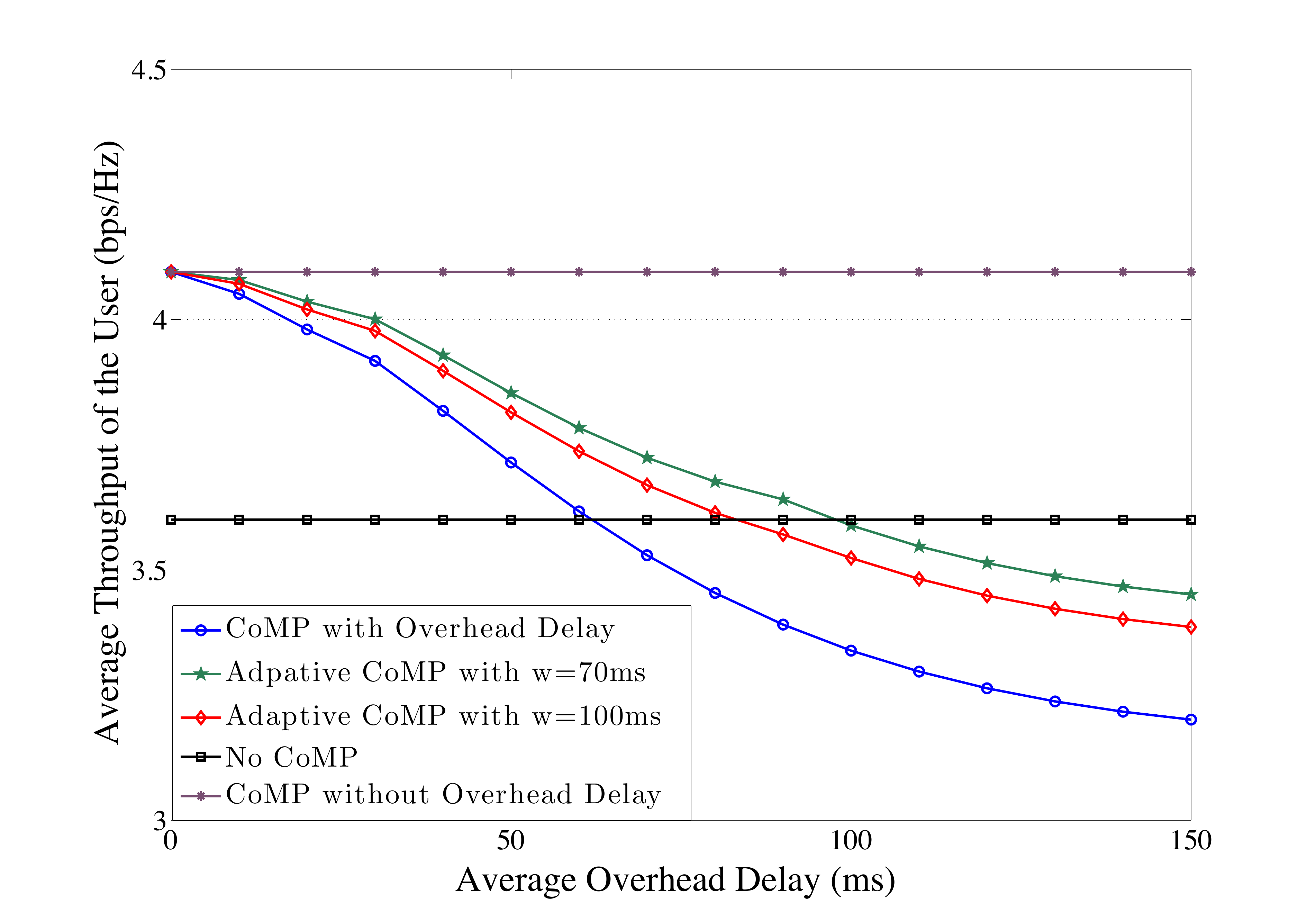}
\caption{User's average throughput vs. the average overhead channel delay for adaptive CoMP ZFBF. The overhead bit size is the same as that used in Fig. \ref{Fig:ThrputCoMP}  and the overhead delay is uniformly distributed in [0, 150ms].}
\label{Fig:ThrputAdaptiveCoMP}
\end{figure}

The simulation results for the adaptive CoMP ZFBF are shown in Fig. \ref{Fig:ThrputAdaptiveCoMP} for two different values of the waiting time duration $w$. Apparently, the adaptive CoMP scheme indeed makes the throughput loss much robust to the  increase of the overhead delay. Namely, it significantly mitigates the impact of the overhead delay on the throughput such that more throughput gain is achieved. Fig. \ref{Fig:ThrputAdaptiveCoMP}  also indicates that the magnitude of the waiting time duration  also affects the performance of the adaptive CoMP scheme. Using an inappropriate $w$ would seriously weaken the performance of the adaptive CoMP scheme. For example, the critical overhead delay for no throughput gain is 82ms as $w=100$ms, which is much shorter than 98ms as $w=70$ms. 

\section{Conclusions}
This paper proposed an adaptive CoMP scheme to mitigate the loss of the throughput gain owing to imperfect overhead messaging. The basic idea of the adaptive CoMP scheme is to eliminate the interference from the coordinated BSs without receiving the updated overhead message within a predesignated waiting time duration. The bounds on the CCDF of the user's SIR with the adaptive CoMP ZFBF are characterized and they are seriously impacted by the waiting time duration. The average throughput of a user for the adaptive CoMP ZFBF scheme was formulated as well. Numerical results showed that the impact of the delayed overhead messages on the user throughput can be significantly reduced by the proposed adaptive CoMP scheme if the waiting time duration is appropriately chosen. 

\appendix
\subsection{Proof of Proposition \ref{Prop:BoundsCCDFSIR}}\label{App:ProofCCDFSIR}
(i) \textbf{The lower bound.} Let $I_n$ denote the user  interference normalized by $p_{k_*}$ and it is given by
\begin{align} \label{EquInterfKtier}
I_n\triangleq \sum_{k=1}^{K}\frac{p_k}{p_{k_*}} \sum_{\mathtt{B}_{i,k} \in \Phi_k \setminus \{ \mathtt{B}_{1,k_*}\}} \delta_{i,k} G_{i,k}|\mathtt{B}_{i,k}|^{-\alpha_k}.
\end{align}
Using \eqref{EquInterfKtier} and Markov's inequality, the upper bound on the CCDF of the SIR can be shown as
\begin{align}\label{UppBoundSIRCDF}
&1-\mathbb{P} \left( \frac{|\mathtt{B}_{1,k_*}|^{\alpha_{k_*}}I_n}{G_{1,k_*}} \leq \frac{1}{\beta}\right)\geq 1-\frac{\beta}{n_{k_*}-|\mathcal{S}_{1,k_*}|-1} \Bigg\{ \nonumber \\ &\sum_{i=2}^{\infty}\mathbb{E}\left[\delta_{i,k_*}\frac{|\mathtt{B}_{1,k_*}|^{\alpha_{k_*}}}{|\mathtt{B}_{i,k_*}|^{\alpha_{k_*}}}\right] 
+ \sum\limits_{\begin{subarray}{c}
k=1\\ k\neq k_*\end{subarray}}^{K}\frac{p_k}{p_{k_*}} \sum_{i=1}^{\infty} \mathbb{E}\left[\delta_{i,k}\frac{|\mathtt{B}_{1,k_*}|^{\alpha_{k_*}}}{|\mathtt{B}_{i,k}|^{\alpha_k}}\right]\Bigg\}.
\end{align}
According to the result of Appendix B in \cite{XiaLiuAndrews13}, we know
\begin{align}\label{Eqn:ithDisRio}
\mathbb{E}\left[\delta_{i,k_*}\frac{|\mathtt{B}_{1,k_*}|^{\alpha_{k_*}}}{|\mathtt{B}_{i,k_*}|^{\alpha_{k_*}}}\right]=\mathbb{E}[\delta_{i,k_*}]\frac{\Gamma\left(1+ \frac{\alpha_{k_*}}{2} \right)(i-1)!}{\Gamma\left(i+\frac{\alpha_{k_*}}{2}\right)}
\end{align} 
for $k\neq k_*$, $\mathtt{B}_{i,k}$ and $\mathtt{B}_{1,k_*}$ are from two independent PPPs. Furthermore, we can show
\begin{align}\label{EquDifferentK}
\mathbb{E}\left[\delta_{i,k}\frac{|\mathtt{B}_{1,k_*}|^{\alpha_{k_*}}}{|\mathtt{B}_{i,k}|^{\alpha_{k}}}\right]
&= \mathbb{E}[\delta_{i,k}]\frac{(\lambda_{k} \pi)^{\frac{\alpha_{k}}{2}}\Gamma\left(1+ \frac{\alpha_{k_*}}{2} \right)(i-1)!}{(\lambda_{k_*} \pi)^{\frac{\alpha_{k_*}}{2}}\Gamma\left(i+\frac{\alpha_k}{2}\right)}
\end{align}
Substituting  \eqref{Eqn:ithDisRio} and \eqref{EquDifferentK} into \eqref{UppBoundSIRCDF} arrives at the lower bound on the CCDF of the SIR in \eqref{Eqn:BoundKtierSIR}.\\
(ii)\textbf{The upper bound.} Since  BSs in different tiers have different powers and path loss exponents, we first have to unify the discrepancies of the powers and path loss exponents in the interference. The interference from the $k$th tier can be expressed as
\begin{align}
I_k & = \sum\limits_{\tilde{\mathtt{B}}_{i,k} \in \hat{\Phi}_k}  p_{k_*} \delta_{i,k}G_{i,k} \left|\tilde{\mathtt{B}}_{i,k}\right|^{-\alpha_k},
\end{align}
where $\tilde{\mathtt{B}}_{i,k} \triangleq \mathtt{B}_{i,k}{\left(p_{k_*}/p_k\right)^{\frac{1}{\alpha_k}}}$. According to the conservation property in \cite{DSWKJM96} , $I_k$ can be viewed as the interference generated from the $K$ new PPPs $\{\tilde{\Phi}_1, \ldots,\tilde{\Phi}_K\}$ with the same transmitting power $p_{k_*}$. Therefore, the normalized interference in (\ref{EquInterfKtier}) can be equivalently rewritten as $I_n \stackrel{d}{=} \sum_{k=1}^{K}I_k$
where $\stackrel{d}{=}$ means equivalence in distribution.

Then replacing all the path loss exponents with the largest one $\alpha_{\max}$ among all path loss exponents, the normalized interference can be lower-bounded by
\begin{align}
I_n \geq \sum\limits_{k=1}^{K} \sum\limits_{\tilde{\mathtt{B}}_{i,k} \in \tilde{\Phi}_k \setminus \{\tilde{\mathtt{B}}_{1,k_*}\}} \delta_{i,k} G_{i,k} |\tilde{\mathtt{B}}_{i,k}|^{-\alpha_{\max}},\,a.s.
\end{align}
Let $\tilde{\Phi}_*= \bigcup_{k=1}^K \tilde{\Phi}_k$ and it is also a PPP with intensity $\tilde{\lambda}_*= \sum_{k=1}^{K} \tilde{\lambda}_k$ since   $\{\tilde{\Phi}_1,\ldots,\tilde{\Phi}_K\}$ are independent PPPs. The lower bound on $I_n$ can be viewed as the interference generated from a single- tier HetNet where the BSs form a PPP $\tilde{\Phi}_*$ with intensity $\tilde{\lambda}_*$ and they have the same path loss exponent $\alpha_{\max}$ and use unit transmit power. Therefore, we can further have
\begin{align}
I_n\geq& \sum_{\tilde{\mathtt{B}}_{i,k} \in \tilde{\Phi}_*} \delta_{i,k} G_{i,k} |\tilde{\mathtt{B}}_{i,k}|^{-\alpha_{\max}},\, a.s.
\end{align}
Then it follows that 
\begin{align*}
\mathbb{P}\left(  \frac{G_{1,k_*} }{I_n|\mathtt{B}_{1,k_*}|^{\alpha_{k_*}}} \geq \beta \right) \leq  \exp\left[-\pi\tilde{\lambda}_*\frac{\Gamma\left(1+\frac{2}{\alpha_{\max}}\right)}{\mathbb{E}[\tilde{Z}])^{\frac{2}{\alpha_{\max}}}}\right], 
\end{align*}
where $\tilde{Z} = \frac{G_{1,k_*}|\mathtt{B}_{1,k_*}|^{-\alpha_{k_*}}}{\beta[3^{-\alpha_{\max}}\delta_{k_*}+(2m+3)^{-\alpha_{\max}}(1-\delta_{k_*})]}$. According to Lemmas 1 and 3 in \cite{JoSanXiaAnd11}, we have $\mathbb{E}[|\mathtt{B}_{1,k_*}|^{-\alpha_{k_*}}]=(\pi \tilde{\lambda}_*)^{-\alpha_{k_*}} \Gamma(1-\alpha_{k_*}/2)$, which gives 
\begin{align}
\mathbb{E} [\tilde{Z}] = \frac{(n_{k_*}-|\mathcal{S}_{1,k_*}|) (\pi \tilde{\lambda}_*)^{\frac{\alpha_{k_*}}{2}} \Gamma \left(1- \frac{\alpha_{k_*}}{2}\right)}{\beta[3^{-\alpha_{\max}}\delta_{k_*}+(2m+3)^{-\alpha_{\max}}(1-\delta_{k_*})]}.
\end{align}
Therefore, the upper bound on \eqref{Eqn:BoundKtierSIR} is obtained. $\mathbb{E}[\delta_{i,k}]$ can be found by directly taking the expectation at the both sides of \eqref{Eqn:delta} and noting that $\mathds{1}(\mathtt{B}_{i,k}\in\mathcal{S}_{1,k_*})$ and $\mathds{1}(\mathtt{B}_{i,k}\notin\mathcal{S}_{1,k_*})$ are deterministic for a given set $\mathcal{S}_{1,k_*}$. 

\subsection{Proof of Proposition \ref{Prop:TimeFrac}}\label{App:TimeFracCOS}
The BS $\mathtt{B}_{i,k}$ is in the COS only when the event $\min\{L_{i,k}, w\}\geq D_{i,k}$ is true, i.e. $\mathtt{B}_{i,k}$ timely receives the updated overhead message within $w$ before another new message is generated. The fraction of the average of the time when  $\mathtt{B}_{i,k}$ is in the COS can be defined as follows
$$\eta_{i,k}\triangleq \lim\limits_{N\rightarrow\infty}\frac{\mathbb{E}\left[\sum^{N}_{j=1}(\min\{L_{i,k}(j),w\}-D_{i,k}(j))^+\right]}{\mathbb{E}\left[\sum_{j=1}^{N}L_{i,k}(j)\right]},$$
where $(a)^+\triangleq\max\{a,0\}$, $L_{i,k}(j)$ and $D_{i,k}(j)$ denote $L_{i,k}$ and $D_{i,k}$ at the $j$th duration of the $N$ time blocks, respectively. Since $L_{i,k}(\cdot)$ and $D_{i,k}(\cdot)$ are i.i.d. at different times, $\eta_{i,k}$ can be simplified as
\begin{align*}
\eta_{i,k}=& \mu_{i,k}\bigg\{\mathbb{E}[\min\{L_{i,k},w\}]-\mathbb{P}[\min\{L_{i,k},w\}\geq D_{i.k}]\cdot\\
&\mathbb{E}[D_{i,k}|\min\{L_{i,k},w\}\geq D_{i.k}]]\bigg\}.
\end{align*}
Also, we know $\mathbb{P}[\min\{L_{i,k},w\}\geq D_{i,k}]=\mathbb{P}[L_{i,k}\geq w]\mathbb{P}[ D_{i,k}\leq w]$ and
\begin{align*}
&\mathbb{E}[\min\{L_{i,k},w\}]= \mathbb{E}[L_{i,k}|L_{i,k}\leq w]\mathbb{P}[L_{i,k}\leq w]\\
&+w\mathbb{P}[L_{i,k}>w]=w-\int_{0}^{w}F_{L_{i,k}}(x)\dif x=\int_{0}^{w} F^{c}_{L_{i,k}}(x)\dif x,
\end{align*}
\begin{align*}
&\mathbb{E}[D_{i,k}|\min\{L_{i,k},w\}\geq D_{i,k}]=\mathbb{E}\bigg[\min\{L_{i,k},w\}\\
& F_{D_{i,k}}(\min\{L_{i,k},w\})-\int_{0}^{\min\{L_{i,k},w\}}F_{D_{i,k}}(x)\dif x\bigg]\\
&=\frac{1}{F_{D_{i,k}}(w)\mathbb{E}[F^c_{L_{i,k}}(D_{i,k})]}\bigg\{F^c_{L_{i,k}}(w)\bigg(wF_{D_{i,k}}(w)\\
&-\int_{0}^{w}F_{D_{i,k}}(x)\dif x\bigg)+F_{L_{i,k}}(w)\mathbb{E}\bigg[L_{i,k}F_{D_{i,k}}(L_{i,k})\\
&-\int_{0}^{L_{i,k}}F_{D_{i,k}}(x)\dif x\bigg]\bigg\}.
\end{align*}
By using the above results,  $\eta_{i,k}$ can be explicitly carried out  as shown in \eqref{Eqn:CoorNatTime}.


\bibliographystyle{ieeetran}
\bibliography{IEEEabrv,Ref_CoMPZFBF}

\end{document}